# Electron and phonon properties of noncentrosymmetric RhGe from *ab initio* calculations

*Maria* Magnitskaya[1,2,*], *Nikolay* Chtchelkatchev[3,1], *Anatoly* Tsvyashchenko[1], *Denis* Salamatin[1], *Sergey* Lepeshkin[2], *Liudmila* Fomicheva[1], and *Mieczysław* Budzyński[4]

[1]Institute for High Pressure Physics, RAS, 108840 Troitsk, Moscow, Russia
[2]Lebedev Physical Institute, RAS, 119991 Moscow, Russia
[3]Landau Institute for Theoretical Physics, RAS, 142432 Chernogolovka, Moscow Region, Russia
[4]Institute of Physics, M. Curie-Skłodowska University, 20-031 Lublin, Poland

**Abstract.** Band structure, Fermi surface, and phonon dispersions of noncentrosymmetric B20-type RhGe are calculated *ab initio* for the first time and their evolution with increasing pressure is investigated. We consider in detail symmetry-conditioned features of the band structure, as well as pressure-induced changes in the Fermi surface topology, which are expected to affect the thermopower of RhGe. We also report on special calculations of electric field gradients on the Rh and Ge nuclei and compare these results with a very recent [111]Cd-TDPAC study of B20-RhGe.

## 1 Introduction

Transition-metal (TM) monosilicides and monogermanides with a noncentrosymmetric B20-type crystal structure attract growing attention due to their special electronic, magnetic, and thermoelectric properties. Thus, FeSi is a magnetic narrow-band semiconductor, while CoSi is a diamagnetic semimetal. The most extensively studied compound MnSi is helimagnetic, as also are MnGe and FeGe. Although TM monogermanides are presently less well investigated than monosilicides, they also exhibit a number of various phenomena. For example, the coexistence of very weak ferromagnetism ($m \sim 7\times10^{-4}$ $\mu_B$/f.u., $T_m = 140$ K) and superconductivity ($T_c = 4.3$ K) has recently been observed in the cubic B20-type high-pressure phase of RhGe with lattice parameter $a = 4.85954$ Å and atomic coordinates $u_{Rh} = 0.12809a$ and $u_{Ge} = 0.83368a$ [1]. Since high-pressure experiments are very time and labor consuming, *ab initio* calculations become a powerful tool for understanding properties of such materials. In this paper, we present the first *ab initio* study of electronic and lattice-dynamical properties of B20-type RhGe (in paramagnetic state), including their evolution on uniform compression. Previous *ab initio* calculations of B20-RhGe published by Tsvyashchenko *et al.* [1], along with their experimental findings, have been devoted to the structure, stability, and possible magnetic arrangement. To our knowledge, except for paper [1], there have been neither experimental nor theoretical studies of B20-RhGe in the literature, so we correlate our results with available information on other B20-type compounds. We also performed *ad hoc* calculations of electric field gradients at the Rh and Ge sites, which allowed us to make a direct comparison with recent [111]Cd time-differential perturbed angular γγ-correlation (TDPAC) measurements of quadrupole frequency in B20-RhGe briefly published in [2].

## 2 Calculation procedure

Our *ab initio* computations are based on the density functional theory (DFT), with the generalized-gradient approximation (GGA) PBEsol for exchange-correlation functional. The lattice-dynamical calculations are conducted within the density-functional perturbation theory (DFPT). We used the first-principles pseudopotential method as implemented in the Quantum Espresso package [3], with the projected-augmented-wave (PAW) type scalar-relativistic pseudopotentials [3]. The valence electron configurations $4s^2p^6d^85s^1$ for Rh and $3d^{10}4s^2p^2$ for Ge were taken. The 24×24×24 **k**-point sampling was employed for integration over the irreducible Brillouin zone (BZ). The plane wave cutoff of 100 Ry was chosen, which gives the total energy convergence of $10^{-7}$ Ry. Structure optimization was performed for all calculated volumes with forces on the ions reduced to less than 3 meV/Å. Phonon dispersions were computed using the interatomic force constants based on a 4×4×4 **q**-point grid, while a 16×16×16 grid was used to obtain the phonon densities of states. Here, we present the results of non-spin-polarized calculations for the paramagnetic state of B20-RhGe.

The listed calculation parameters allowed us to obtain the equilibrium specific volume $\Omega_0$ that is only 0.5% smaller the experimental value at $p \approx 0$ GPa. On the whole, our results on the structural parameters and stability of B20-RhGe are consistent with experimental and theoretical data reported in paper [1].

---
[*] Corresponding author: magnma@gmail.com

## 3 Results and discussion

### 3.1. Band structure and Fermi surface

The simple cubic FeSi-type structure, usually referred to as B20 (space group #198, P2$_1$3) is described in detail, e.g., in paper [1]. As has been comprehensively explained with an example of B20-MnSi [4], the P2$_1$3 structure gives rise to uncommon electronic properties due to its noncentrosymmetric and nonsymmorphic nature. Figs. 1a and 1b display correspondingly the electronic band structure along high-symmetry directions of BZ and the Fermi surface (FS) at a reduced volume of $0.95\Omega_0$ (~8 GPa). Actually, the bands computed at other pressures (0, 22, and 43 GPa) look very similar, except the neighborhood of the $\Gamma$ point. The overall shape of the RhGe band structure is much like those previously calculated for other nonmagnetic B20-type TM monosilicides [4–7] and monogermanides [8, 9]. The same is true for the RhGe density of states (DOS) (not depicted) which, in the energy range of interest ( –4 eV to +2 eV), is contributed mostly by the hybridized Rh 4d- and Ge 3p-states, with a dominating contribution from the former. A rather low value of DOS at the Fermi level ( ~2 states/eV per cell) implies that cubic RhGe is possibly a semimetal or poor metal.

Noteworthy are some peculiarities of the band structure: all the bands along the path R–X–M–R corresponding to the BZ face are doubly degenerate (and relatively low-dispersive) and all the states at the zone corner R are four-fold degenerate. Another uncommon feature is the presence of triply degenerate levels at the zone center $\Gamma$ ($k = 0$), one of which falls exactly at $E_F$ if RhGe is compressed to $\Omega = 0.9\Omega_0$ ( ~22 GPa). The involved bands marked out by rectangular figure are located along the M–$\Gamma$–R path around $E_F$. One of the three degenerate levels has a zero slope (flat band) and the other two have $k$-linear dispersions of opposite sign, similar to the Dirac cone in graphene. The same feature situated almost exactly at $E_F$ is observed in isoelectronic compounds CoSi, RhSi, and CoGe [5, 6, 8, 9].

According to the rigid band approximation, the Dirac point is expected to lie above $E_F$ for non-isoelectronic B20 compounds with one missing electron per unit cell. This is confirmed by *ab initio* calculations of MnSi [4] and other 3d- and 4d-metal monosilicides [5–7]. The degenerate and low-dispersive bands manifest themselves in the interband transitions, which can be observed in the measurements of optical spectra.

We explored the variation with pressure of the band structure and Fermi surface. The evolution of the Dirac point upon compression is as follows. At normal pressure $p \approx 0$ GPa, it lies slightly below $E_F$. On compression, it goes up, crosses the $E_F$ at 22 GPa and then increases further. Accordingly, the initial $p = 0$ Fermi surface consists of an electron pocket at the zone center $\Gamma$ related to the Dirac point, electron pockets at the BZ corner R, and hole pockets at the M point (center of the BZ edge). All the pockets are rather small and the FS resembles that of a semimetal. At 8 GPa, a cage-like structure of holes appears which is associated with the flat band that crosses the Fermi level near the $\Gamma$-point. Interestingly, the same cage-like structure was obtained for the normal-pressure FS of CoGe [9], however, the authors were not sure about this structure, considering its presence to be beyond the accuracy of their calculations. On further compression, the cage-like structure continues to develop, mainly in the $\Gamma$–R and $\Gamma$–M directions, and at 43 GPa it eventually becomes a closed simply-connected surface. The Dirac-point electron pocket centered on $\Gamma$ vanishes at 22 GPa. This occurrence, as well as appearance of the hole cage-like structure at 8 GPa and change of its connectivity at 43 GPa are indicative of pressure-induced electronic topological transitions (ETT).

One way of experimentally identifying an ETT is to measure the electrical resistivity and thermopower which commonly have peculiarities at the ETT. Worthy of mention are the combined theoretical and experimental studies of non-magnetic CoSi [5] and CoGe [8] which are considered as promising metallic thermoelectrics. A low resistivity and a large negative Seebeck coefficient $S$ are measured for both CoSi and CoGe. Upon hole doping, a large positive $S$ is observed. The authors [8] demonstrate that the Dirac-point feature is the origin for the large magnitude of positive or negative $S$. Turning back to RhGe, we see that the applied pressure acts in the same direction as the hole doping: the Dirac-point feature goes up relative to $E_F$. Based on this observation, we expect that similar pressure-induced thermopower anomalies may exist in B20-RhGe.

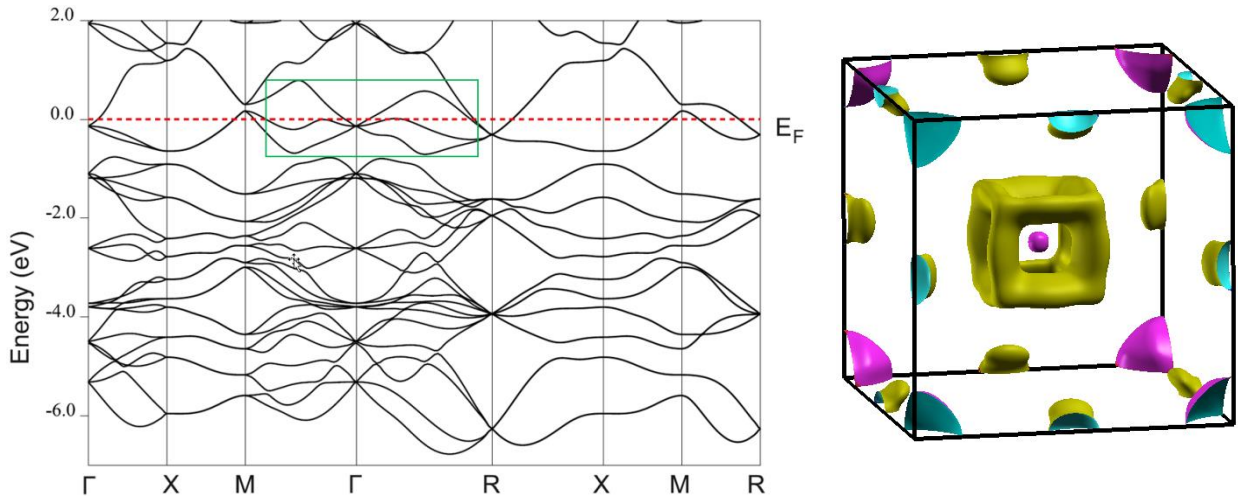

**Fig. 1.** (color online). (a) The band structure diagram of B20-RhGe at normal pressure, energy is measured from the Fermi level $E_F$ (dashed line). The Dirac-point feature is enclosed in the rectangle (see text). (b) The Fermi surface at 8 GPa. Small electron pocket in the BZ center ($\Gamma$ point) corresponds to the Dirac-point feature. The evolution with pressure of the electron-like (pink online) and hole-like (green online) sheets of FS is described in the text.

### 3.2. Lattice dynamics

The phonon dispersions and the phonon density of states (PDOS) of B20-RhGe are shown in Fig. 2. To our knowledge, there are neither theoretical nor experimental studies of the B20-RhGe phonon properties, so we compare our results with the available lattice-dynamical calculations of $M$Si ($M$ = Fe, Ru, Os) [7] and Fe$_{1-x}$Co$_x$Si [10], the last system was also studied by the inelastic neutron scattering. The general form of the phonon dispersions and PDOS is similar to those obtained in [7, 10]. Again, as in the case of electron energy bands, the phonon modes along the face R–X–M–R are two-fold degenerate. The evolution of PDOS with pressure is demonstrated consecutively in Fig. 2b from top to bottom. At zero pressure, there is a gap between the optical low-frequency and high-frequency modes, which disappears on compression.

Noteworthy is the pressure behavior of transverse acoustic modes TA1 and TA2 propagating along the [110] (ΓM) direction. It is known that in cubic crystals, the TA1 and TA2 modes are different only in the case of propagation in the [110] direction. With increasing pressure, these modes approach each other and practically merge at 30 GPa, as shown in Fig. 2a. This fact should also manifest itself in the behavior of elastic constants. In the [110] direction, the TA1 and TA2 sound velocities $v$ are expressed through the shear constants $C\,' = (C_{11} - C_{12})/2$ and $C_{44}$, respectively: $v_{T1} = (C\,'/\rho)^{1/2}$ and $v_{T2} = (C_{44}/\rho)^{1/2}$, where $\rho$ is the density. For the lack of information on the phonon spectra of B20 compounds under pressure, we compare our results with the study of isovalent B20-RhSi, whose elastic constants were evaluated up to 35 GPa [11]. There is a qualitative correlation in behavior between the RhGe TA phonon modes in ΓM direction (Fig. 2a) and the RhSi elastic constants $C\,'$ and $C_{44}$: as pressure increases from 0 to 30 GPa, the acoustic shear anisotropy $(C_{44} - C\,')/C_{44}$ decreases from 0.24 to 0.03 [11]. Hopefully, our lattice-dynamical calculations may be useful as a reference for future studies of RhGe, e.g. neutron scattering measurements.

### 3.3. The electric field gradients

In addition, we performed special calculations of the electric field gradient (EFG) created on a lattice site by electronic and ionic environment. These results were directly compared with experimental data on the quadrupole precession frequency, which has been measured [2] by $^{111}$Cd-TDPAC method. The experimental technique is described in [12]. The EFG is defined as the second derivate of the electric potential $V$ at a nucleus. The principal-axis components of the EFG tensor $V_{ii} = \partial^2 V/\partial i^2$. In cubic crystals, the EFG tensor is reduced to the only component $V_{zz}$. The EFG is expressed as $V_{zz} = h\nu_Q/eQ$, where $Q$ is the nucleus quadrupole moment and $\nu_Q$ is the quadrupole precession frequency.

These *ad hoc* calculations for paramagnetic RhGe were carried out using the APW+lo method developed in the Wien2k package [13]. This approach is all-electron, so it is quite suitable for exploring the immediate vicinity of a nucleus, where the EFG is significant ($V_{zz}$ scales as $1/r^3$). We briefly list the calculation parameters: the MT-radii $R_{Rh}$ = 2.42 a.u. and $R_{Ge}$ = 2.30 a.u., the energy cutoff $R_{min}K_{max}$ = 9.0, the maximal number of **k** points was 1135. The total energy was converged to $10^{-8}$ Ry and the residual forces on the ions were reduced to 10 meV/Å. As is known, the EFG is non-zero only if the charge density that surrounds the nucleus violates cubic symmetry and therefore generates an inhomogeneous electric field. It is the case of P2$_1$3 structure: although it belongs to cubic system, its overall point symmetry is tetrahedral $T^4$ rather than cubic $O_h$ and the local point symmetry at the Rh and Ge sites is rhombohedral (R3).

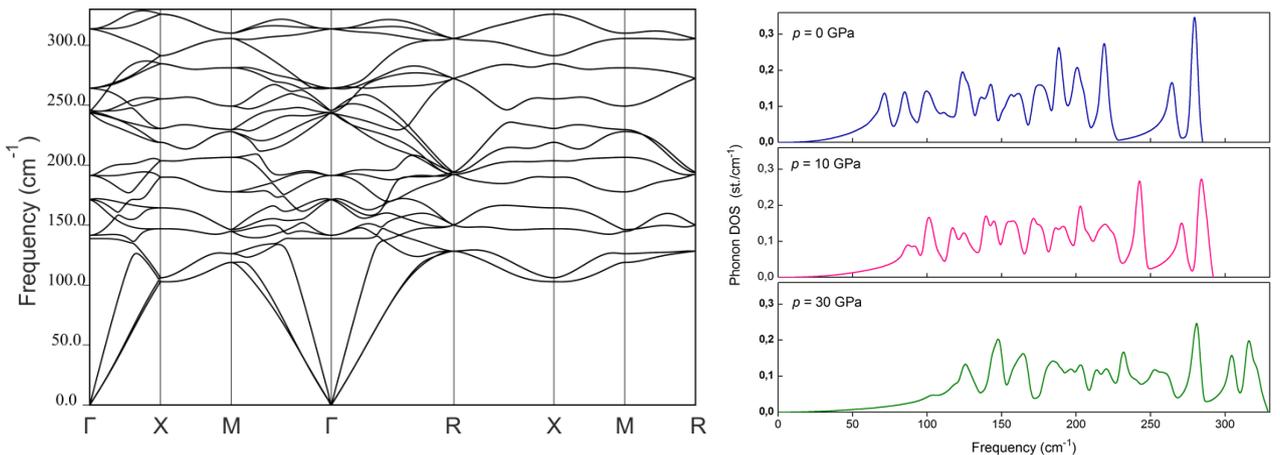

**Fig. 2.** (color online). (Left) Phonon dispersions of B20-RhGe at 30 GPa. (Right) From top to bottom: the phonon density of states at 0, 10, and 30 GPa.

---
\* Corresponding author: magnma@gmail.com

At the experimental volume $\Omega_0$, our ground-state ($T = 0$ K) calculations give $V_{zz}^{Rh} = 1.7 \times 10^{17}$ V/cm$^2$ for the EFG on the Rh site. The $V_{zz}^{Ge} = -7.4 \times 10^{17}$ V/cm$^2$ is several times larger in magnitude and opposite in sign. Experimentally, in the paramagnetic region ($T > 140$ K), the frequency $\nu_Q$ measured on the $^{111}$Cd probe nuclei introduced to the RhGe lattice is equal to 134.1 MHz at $T_{room}$ and 134.6 MHz at 160 K [2]. The corresponding $V_{zz}$ values are $6.68 \times 10^{17}$ V/cm$^2$ and $6.71 \times 10^{17}$ V/cm$^2$ (the TDPAC experiments on polycrystalline samples do not provide information on the EFG sign). Thus, experimental EFG on the $^{111}$Cd nucleus, $V_{zz}^{Cd}$, significantly exceeds in magnitude the $V_{zz}^{Rh}$ and differs by only 10% from $V_{zz}^{Ge}$. This suggests that $^{111}$Cd probes substitute for the Ge atoms in RhGe and therefore provide information on the Ge sublattice. It is common that the experimental EFG weakly increases with decreasing temperature. However, even a hypothetical zero-temperature value of $V_{zz}^{Cd}$ would hardly reach the calculated $V_{zz}^{Ge}$, since the non-sphericity of electron density on the Ge atom with its directed p-orbitals is much more pronounced than on the Cd atom with filled s- and d-shells. The results of our study of electric field gradients in B20-RhGe will be published in detail elsewhere.

## 4 Conclusions

To conclude, *ab initio* density-functional calculations are conducted of the B20-type noncentrosymmetric high-pressure phase of RhGe. We explored the electronic band structure, Fermi surface, and phonon dispersions in paramagnetic RhGe under normal and high pressure. For the lack of theoretical and experimental data on the electron and phonon properties of B20-RhGe in the literature, we correlated the obtained results with available information on other B20-type TM monosilicides and monogermanides. On the whole, our results are consistent with previous calculations of isovalent nonmagnetic B20 compounds. The symmetry-conditioned Dirac-point feature of the B20-type band structure is demonstrated to be essential for the FS evolution with pressure. We consider the pressure-induced changes in the FS topology, which are expected to affect the thermopower of RhGe.

In addition, we performed special *ab initio* calculations of the electric field gradients on the Rh and Ge sites and compared the results with recent $^{111}$Cd-TDPAC measurements of quadrupole frequency in the RhGe lattice. The comparison allows us to suggest that the $^{111}$Cd probes occupy the metalloid sites in the RhGe lattice and therefore, provide information on the local environment of Ge atoms.

---

\* Corresponding author: magnma@gmail.com

This work is supported by Russian Academy of Sciences and Russian Foundation for Basic Research (grants 16-02-01122 and 16-32-00922). The numerical calculations are performed using computing resources of the Joint Supercomputer Center of RAS and the federal collective usage center Complex for Simulation and Data Processing for Mega-science Facilities at NRC "Kurchatov Institute", http://ckp.nrcki.ru/. A.T. and D.S. acknowledge the support from Russian Scientific Foundation (grant RNF 17-02-01050).